\documentclass[prd,aps,twocolumn,superscriptaddress,floatfix,preprintnumbers]{revtex4-1}
\usepackage{amsmath}
\usepackage{graphicx}
\usepackage{xcolor}
\usepackage{amsfonts}
\usepackage{amssymb,amsmath,bm}
\usepackage[citecolor=blue,pdfa=true,linktocpage=true,urlcolor=blue,colorlinks=true]{hyperref}
\usepackage[export]{adjustbox}
\usepackage{multirow}
\usepackage{physics}
\usepackage{cleveref}
\usepackage{soul}

\newcommand{\aap}{{Astron.~Astrophys.}}
\newcommand{\apjl}{{Astrophys.~J.~Lett.}}

\newcommand{\mnras}{{Mon.~Not.~R.~Astron.~Soc.}}
\newcommand{\jcap}{JCAP}

\newcommand{\be}{\begin{equation}}
\newcommand{\ee}{\end{equation}}

\newcommand{\bea}{\begin{eqnarray}}
\newcommand{\eea}{\end{eqnarray}}
\newcommand{\bdm}{\begin{displaymath}}
\newcommand{\edm}{\end{displaymath}}

\newcommand{\vk}{{\bf k}}

\def\fNL{f_{\mathrm{NL}}}

\newcommand{\bk}{\mathbf{k}}

\def\bx{\bold{x}}

\newcommand{\hr}[1]{{\color{black}#1}}

\begin{document}

\title{Exact Modeling of Power Spectrum Multipole through Spherical Fourier-Bessel Basis}

\author{Robin Y. Wen}
\email{ywen@caltech.edu}

\affiliation{California Institute of Technology, 1200 East California Boulevard, Pasadena, CA 91125, USA}

\author{Henry S. Grasshorn Gebhardt}
\affiliation{California Institute of Technology, 1200 East California Boulevard, Pasadena, CA 91125, USA}
\affiliation{Jet Propulsion Laboratory, California Institute of Technology, Pasadena, California 91109, USA}

\author{Chen Heinrich}
\affiliation{California Institute of Technology, 1200 East California Boulevard, Pasadena, CA 91125, USA}

\author{Olivier Dor\'e}
\affiliation{California Institute of Technology, 1200 East California Boulevard, Pasadena, CA 91125, USA}
\affiliation{Jet Propulsion Laboratory, California Institute of Technology, Pasadena, California 91109, USA}

\begin{abstract}
 
The three-dimensional galaxy power spectrum is a powerful probe of primordial non-Gaussianity and additional general relativistic effects, which become important on large scales. At the same time, wide-angle (WA) effects due to differing lines-of-sight (LOS) on the curved sky also become important with large angular separation. In this work, we accurately model WA and Doppler effects using the spherical Fourier-Bessel (SFB) formalism, before transforming the result into the commonly used power spectrum multipoles (PSM). This mapping from the SFB power spectrum to PSM represents a new way to non-perturbatively model WA and GR effects present in the PSM, which we validate with log-normal mocks. Moreover, for the first time, we can compute the analytical PSM Gaussian covariance on large scales, exactly including WA-induced mode-couplings, without resorting to any plane-parallel approximations. 
\end{abstract}
\maketitle

\section{Introduction}
Current and upcoming large-scale structure (LSS) surveys such as DESI~\cite{16DESI}, Euclid~\cite{16Euclid}, and SPHEREx~\cite{14SPHEREx} will measure the galaxy density field over increasingly larger volume, offering improved constraining power on effects that manifest on large scales, such as the primordial non-Gaussianity (PNG)~\cite{01Komatsu_fNL,07Dalal_PNG,08Slosar_fnl,17dePutter_fnl1,23_DESI_LRG} and general relativistic (GR) effects~\cite{09Yoo_GR,10Yoo_GR,11ChallinorLPS,11Bonvin_GR,12JeongLPS,22Elkhashab_Euclid_like}. With wider angular coverage, the commonly used global plane-parallel approximation, which assumes the same line-of-sight (LOS) for all galaxy pairs, breaks down when the galaxy separation becomes large, and wide-angle (WA) effects become important. Additionally, the Newtonian 
approximation to redshift space distortion (RSD)-- an effect that arises from our measuring the redshifts and not the distances of galaxies -- also breaks down as GR effects become important on large scales. 

The next best thing to the global plane-parallel approximation is a local one in which each galaxy pair is allowed to have a different LOS. This is achieved by the Yamamoto estimator for measuring the Fourier-space power spectrum multipoles (PSM). An efficient implementation of this is to choose one of the galaxy to be the pair LOS, called the end-point LOS~\cite{15Bianchi_FFT,15Scoccimarro_FKP,19BeutlerMutipolePk}. The signal picked up by this estimator at large scales includes WA effects that are usually modeled either perturbatively as an expansion in the galaxy pair separation~\cite{16ReimbergWA,18CastorinaWA,19BeutlerMutipolePk,24Benabou_WA,21BeutlerWAWindow,23WAGR,23PaulWA}, or non-perturbatively via an exact calculation of the configuration-space correlation function \cite{18Tansella-GR,22CatorinaGR-P}. In a companion paper, Ref.~\cite{24Benabou_WA} demonstrates that the perturbative method for WA effects breaks down at large scales, which can bias constraints on PNG and motivate further study of non-perturbative WA modeling.

In this work, we propose an alternative non-perturbative method for calculating the PSM that employs a natural basis for the curved sky: the spherical Fourier-Bessel (SFB) basis, which was proposed in the 90s \cite{91Binney_GaussianRF,93lahav_spherical} and has recently gained traction \cite{12Leistedt,13Yoo_GR_SFB,14Nicola_SFB,15Lanusse_SFB,19Samushia_SFB,20Wang_Hybrid_Estimator,21Zhang_SFB_TSH,21Gebhardt_SuperFab,22Khek_SFB_fast,23Gebhard_SFB_eBOSS}. 
As the eigenfunctions of the Laplacian in the spherical coordinates, the SFB basis preserves the geometry of the curved sky and naturally models the WA effects. Due to the separation of radial and angular modes in spherical coordinates, one can incorporate Newtonian RSD and GR effects more easily for the SFB power spectrum \cite{13Yoo_GR_SFB,21Zhang_SFB_TSH} than for the PSM \cite{22CatorinaGR-P}.

Despite these advantages, the SFB power spectrum is a less developed formalism compared to the PSM, especially at the mildly non-linear scales ($k\gtrsim0.05 \,h/{\rm Mpc}$). Its estimator also suffers from an increased computational cost due to the large number of modes, and the fact that one cannot employ fast Fourier-transforms as is often used for PSM~\cite{21Gebhardt_SuperFab,23Gebhard_SFB_eBOSS}. 

In this work, we build upon Ref.~\cite{18CastorinaWA} and fully develop a mapping from the SFB power spectrum to the PSM in Sec.~\ref{sec:PSM}. This mapping allows us to take advantage of the strengths in both statistics: on large scales we can exactly model WA and GR effects using the SFB power spectrum, before transforming the result into the PSM, to take advantage of its efficient Yamamoto estimator implementation and better developed small-scale modeling. This mapping represents a new way of non-perturbatively modeling WA effects in the PSM, and in Sec.~\ref{sec:Validation}, we validate this new approach to WA modeling using the estimator and log-normal mocks described in a companion paper, Ref.~\cite{24Benabou_WA}. Moreover, for the first time we obtain an exact analytic covariance for the PSM (Sec.~\ref{sec:covariance}), handling off-diagonal mode-couplings due to WA effects in the covariance, without resorting to approximations such as those made in Ref.~\cite{20Wadekar_covariance}. Though formulated in the context of galaxy surveys, our results will also have implications for future wide-field intensity mapping surveys \cite{13Hall_GR,16SFB_IM,20SKA,21Viljoen_Cl_PNG_GR}.

\section{Power Spectrum Multipoles}\label{sec:PSM}

We first briefly review the SFB formalism. The canonical SFB basis is composed of eigenfunctions of the Laplacian in spherical coordinates, namely the spherical Bessel functions of first kind $j_\ell(kx)$ and the spherical harmonics $Y^*_{\ell m}(\hat{\bx})$. The SFB decomposition of the galaxy density field $\delta(\bx)$ is then:
\begin{align}
\delta_{\ell m}(k) &=
\int_{\bx}
j_\ell(kx)Y^*_{\ell m}(\hat{\bx})
\delta(\bx)\,.
\label{eq:sfb_k-to-x}
\end{align}
Here, the angular mode $Y_{\ell m}(\hat{\bx})$ and the radial mode $j_\ell(kx)$ share the same $\ell$, whereas the generalized SFB (gSFB)  decomposition~\cite{18CastorinaWA} allows for different $L\neq\ell$:
\begin{align}
\delta_{\ell m}^{L}(k) &=
\int_{\bx}
j_L(kx)Y^*_{\ell m}(\hat{\bx})
\delta(\bx)\,.
\label{eq:gsfb_k-to-x}
\end{align}
They form an over-complete basis, but as we shall see, provide a useful bridge for expressing the PSM in the SFB formalism. 

The generalized SFB power spectrum is defined as:
\begin{align}
   \langle\delta_{\ell_1 m_1}^{a}(k_1) \delta_{\ell_2 m_2}^{b*}(k_2)\rangle = C_{\ell_1}^{ab}(k_1,k_2)\delta_{\ell_1\ell_2}^{K}\delta_{m_1m_2}^{K}
   \label{eq:gSFB-def}\,,
\end{align}
where $\delta^{K}$ is the Dirac-delta. We use the upper indices to indicate the radial modes and the lower index for the angular mode of the gSFB power spectrum. When $a=b=\ell_1$, it reduces to the canonical SFB power spectrum $C_{\ell_1}(k_1,k_2)$. Note that the SFB power spectra adopt two values of $k$, and we call $k_1=k_2$ diagonal and $k_1\neq k_2$ off-diagonal.

In a real survey, the observed field $\delta^{{\rm W}}$ contains the survey window function: $\delta^{{\rm W}}(\bx)=W(\bx)\delta(\bx)$. This breaks the azimuthal symmetry, but we can define the azimuthally averaged quantity
\begin{align}
C_{\ell}^{ab,{\rm W}}(k_1,k_2)\equiv\frac{1}{2\ell+1}\sum_{m}\langle\delta_{\ell m}^{a,{\rm W}}(k_1) \delta_{\ell m}^{b*,{\rm W}}(k_2)\rangle\,,\label{eq:gSFB-W}
\end{align}
which is now similar in form to Eq.~\ref{eq:gSFB-def}.

Now the power spectrum multipoles (PSM) $P_{L}(k)$ are defined as, under the end-point LOS~\cite{15Scoccimarro_FKP}
\begin{align}
P_{L}(k)&\equiv \frac{(2L+1)}{I_{22}}\int_{\hat{\bk}} \int_{\bx,\bx'} e^{-i \bk \cdot (\bx-\bx')} \nonumber\\
&\qquad \langle \delta(\bx)\delta(\bx') \rangle W(\bx)W(\bx') \mathcal{L}_L(\hat{\bx}\cdot \hat{\bk})\,,
\label{eq:Pl_average}
\end{align}
where $\mathcal{L}_L(\hat{\bx}\cdot \hat{\bk})$ is the Legendre polynomial and $I_{22}\equiv\int_{\bx} W(\bx)^2$ is the normalization. Using the plane-wave expansion  and evaluating the $\hat{\bk}$ integral (see full derivation in the Appendix.~\ref{sec:PSM-SFB_Deriv}), we find the relationship between the PSM and the gSFB power spectrum:
\begin{align}
P_{L}(k)&=\frac{(4\pi)^2(2L+1)}{I_{22}}\sum_{a,b}i^{-a+b}(2a+1)(2b+1)\nonumber\\
&\qquad \begin{pmatrix}
a & L & b\\
0 & 0 & 0
\end{pmatrix}^2 C_{b}^{ab,{\rm W}}(k,k)\,.
\label{eq:PSM-SFB}
\end{align}

The above equation was first derived in Ref.~\cite{18CastorinaWA}. Compared to their expression, our Eq.~\ref{eq:PSM-SFB} accounts for the window convolution and is expressed in terms of the gSFB power spectrum (Eq.~\ref{eq:gSFB-W}) instead of the gSFB mode (Eq.~\ref{eq:gsfb_k-to-x}). Going beyond Ref.~\cite{18CastorinaWA}, we will highlight the theoretical importance of Eq.~\ref{eq:PSM-SFB} and, for the first time, enable its use for theoretical modeling of PSM by developing a method to compute the gSFB power spectrum.

Remarkably, the PS monopole is simply
\begin{align}
P_{0}(k)&=\frac{(4\pi)^2}{I_{22}}\sum_{b}(2b+1) C_{b}^{{\rm W}}(k,k)\,,\label{eq:Pk-0}
\end{align}
a sum over all the angular modes of the canonical SFB power spectrum. Note that only the diagonal components of the canonical SFB power spectrum are present. These are the only modes that would exist in a homogeneous and isotropic Universe where there are no redshift evolution and LOS-dependent effects like RSD~\cite{21Gebhardt_SuperFab}. This is exactly what the PS monopole is, an average over the orientation of the Fourier mode $\vk$, and it erases the redshift evolution information by integrating over the redshift bin.

Given the above, we would then expect that the off-diagonal components $C_b^{\rm W}(k,k')$ would be (partially) brought back for the higher multipoles in $P_L$. Indeed, this is achieved through the generalized SFB $C_b^{ab,\rm W}(k, k)$, where by construction, the off-diagonal information in $C_b^{\rm W}(k,k')$ is folded into the extra upper radial indices $a$ and $b$, keeping a single $k$ value. 

The double sum in Eq.~\ref{eq:PSM-SFB} is also weighted by the Wigner-3$j$ symbols, a result of the PSM being a weighted average over the orientation of $\vk$ by the Legendre polynomials. The resulting triangle condition between $a, b$ and $L$ means that, for the commonly-considered PSM $L = 0$ to 4, Eq.~\ref{eq:PSM-SFB} is effectively reduced to only to a single sum over $b$ (at most a couple hundred of terms for $k\leq 0.1\,h/{\rm Mpc}$) with few terms in $a$, making it computationally efficient. Since WA terms are naturally contained in the gSFB power spectrum, the mapping in Eq.~\ref{eq:PSM-SFB} will offer an exact treatment of WA effects in PSM if we can exactly compute the gSFB.

To compute the window-convolved gSFB power spectrum, we first consider the simplest case of a full-sky window with only radial selection $R(x)$. We use the superscript R to indicate this case. Following Refs.~\cite{21Gebhardt_SuperFab,22Khek_SFB_fast,23Gebhard_SFB_eBOSS}, the radially-convolved gSFB power spectrum can be written as:
\be
C_{\ell}^{ab,{\rm R}}(k_1,k_2)\equiv\int_{q}\mathcal{W}_{\ell}^{a}(k_1,q)\mathcal{W}_{\ell}^{b}(k_2,q)P_{\rm m,0}(q)\,,
\label{eq:gSFB-kernel_form}
\ee
where $P_{\rm m,0}(q)$ is the present-day matter power spectrum, $\mathcal{W}_{\ell}^{a}(k,q)$ is the gSFB kernel defined as
\be
\mathcal{W}_{\ell}^{a}(k,q)\equiv \sqrt{\frac{2}{\pi}}q\int_{x} x^2 j_a(kx)R(x)\Delta_{\ell}(q,x)\,,
\label{eq:gSFB-kernel}
\ee
and $\Delta_{\ell}(k,x(z))$ is the angular kernel that would contain effects such as RSD, GR \cite{11Bonvin_GR,11ChallinorLPS,13DiDio_classgal} and Fingers-of-god (FoG) effects~\cite{22Khek_SFB_fast}. On linear scales where the FoG effect is small, and in the Newtonian limit, the angular kernel becomes
\be
\Delta_{\ell}^{\rm RSD}(k,x)=D(x)\left[b_{\rm g}(x)j_{\ell}(kx)-f(x)j_{\ell}''(kx)\right]\,,
\label{eq:kernel-RSD}
\ee
where $D$ is the growth factor, $b_g$ is the linear galaxy bias, and $f$ is the linear growth rate. 

\begin{figure*}[ht]
\centering
\includegraphics[width=0.9\textwidth]{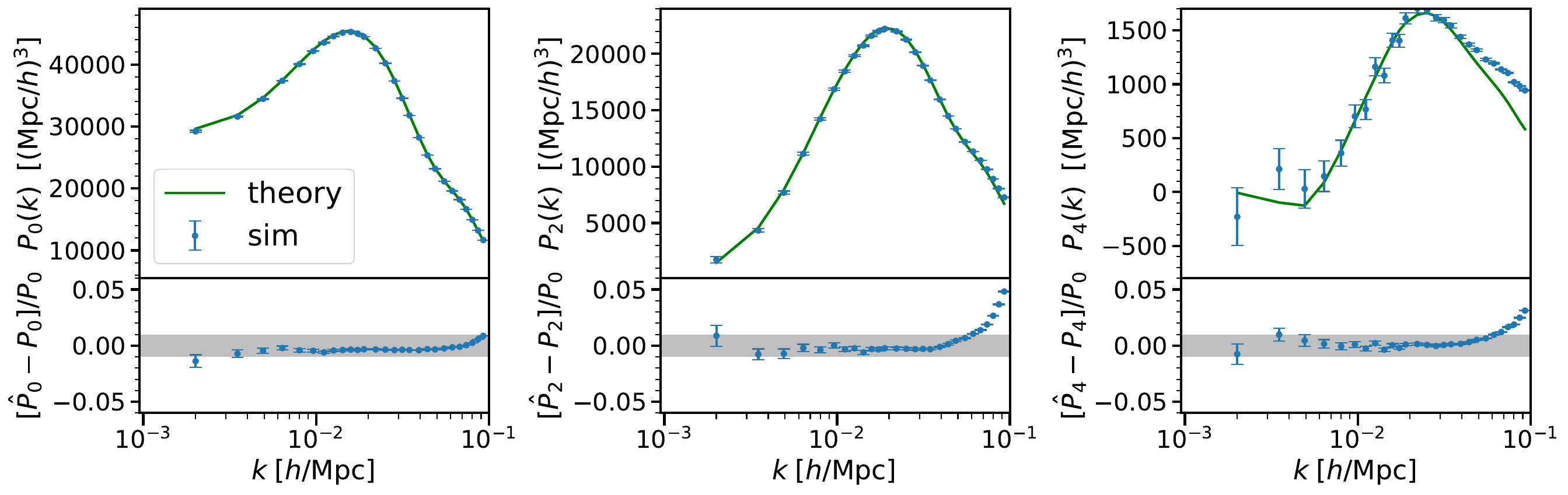}
\includegraphics[width=0.6\textwidth]{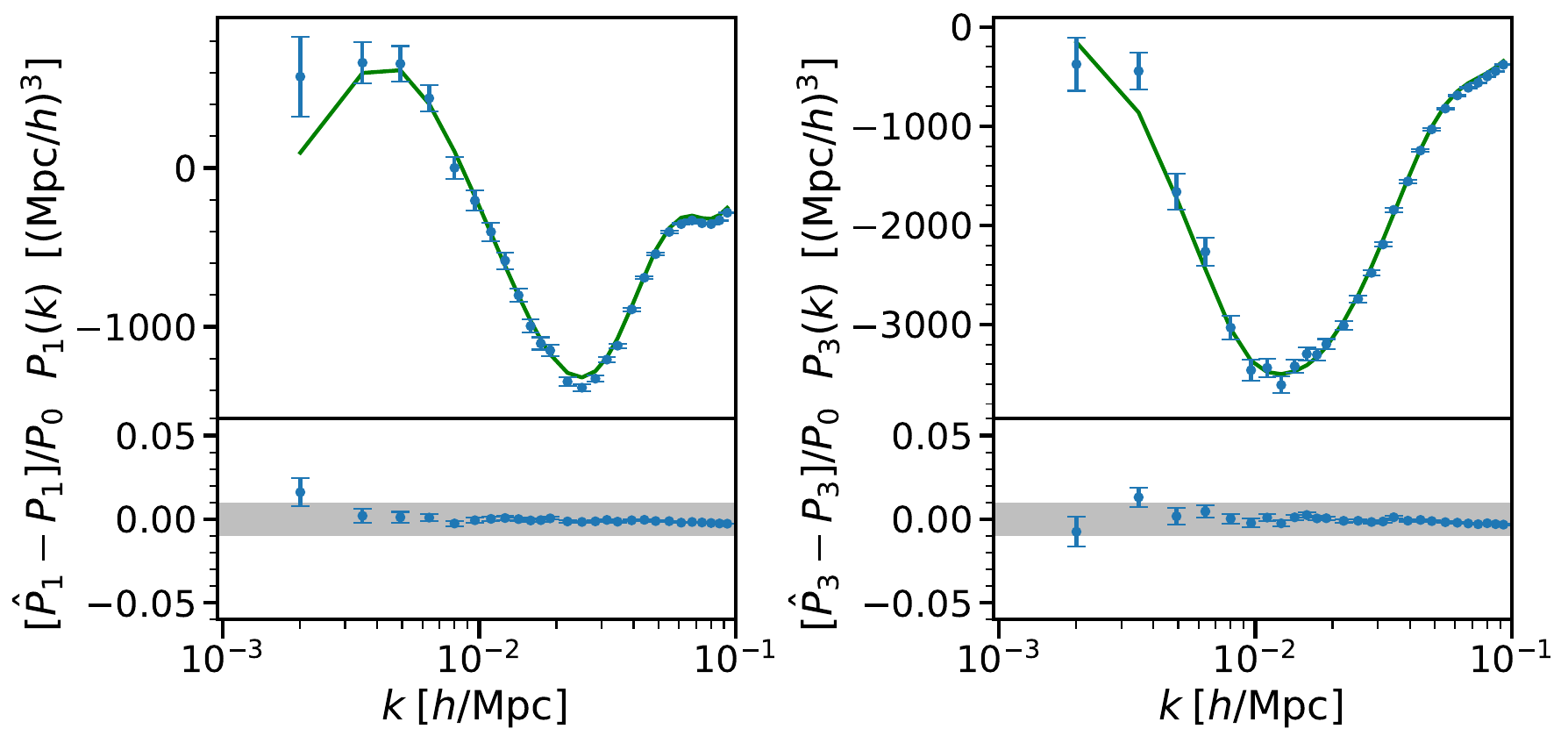}
\caption{The theoretical and measured multipoles $P_{L}(k)$. The green lines show the theoretical results of PSM, and the blue points show the mean and the $1\sigma$ error of the measured multipoles from the 10,000 full-sky mocks for $z=0.2-0.5$ based on the Yamamoto estimator. The $1\sigma$ errors shown here represent the error on the mean instead of the error for the measured spectrum from a single realization, and the errors are obtained by calculating the standard deviation of the 10,000 measured spectra. In the lower panels, we show the differences between the measurements and the theoretical results, normalized by the monopole that sets the simulation precision. The grey bands indicate $1\%$ accuracy. Our theoretical results and measurements agree well at large scales.}
\label{fig:multipole-sims} 
\end{figure*}

The integrals in Eq.~\ref{eq:gSFB-kernel_form} and \ref{eq:gSFB-kernel} can be computed through the procedure introduced in Refs.~\cite{22Khek_SFB_fast} and \cite{23Gebhard_SFB_eBOSS} where the authors computed the canonical SFB power spectrum. The only change we make is allowing radial and angular modes to have different orders $\ell\neq a$ when computing the gSFB. With this we evaluate the radially-convolved gSFB power spectrum and obtain the PSM using Eq.~\ref{eq:PSM-SFB} for an example of the full-sky window (taking $z$ between 0.2 and 0.5). We plot the theory computation for the first five multipoles in solid lines in Fig.~\ref{fig:multipole-sims}.

Going beyond the full-sky window, we assume the separability of the window function into angular and radial parts $W(\bx)=R(x)M(\hat{\bx})$. We leave the radial selection $R(x)$ in the gSFB kernel and consider the effects of the angular mask $M(\hat{\bx})$. The window-convolved gSFB power spectrum is then:
\begin{align}
C_{\ell}^{ab,{\rm W}}(k_1,k_2)=\sum_{\ell'}M_{\ell\ell'}^{}C_{\ell'}^{ab,{\rm R}}(k_1,k_2),
\label{eq:angular-mask-gSFB}
\end{align}
where $M_{\ell\ell'}$ is the angular mode coupling matrix~\cite{19Alonso_Nmaster}. This step enables the use of Eq.~\ref{eq:PSM-SFB} for realistic surveys.

\section{Validation with Full-sky Mocks}\label{sec:Validation}

To validate Eq.~\ref{eq:PSM-SFB} and showcase its use for modeling WA effects, we compare our theoretical results with measurements from full-sky simulations. In particular, we choose to use the set of log-normal mocks generated in Ref.~\cite{24Benabou_WA}. With the goal of studying a perturbative calculation of WA effects, the authors of Ref.~\cite{24Benabou_WA} generated 10,000 full-sky mocks in linear theory with Newtonian RSD, spanning $z=0.2$ to $0.5$. They also provide a new implementation for the following Yamamoto estimator, whose ensemble average is expected to be Eq.~\ref{eq:Pl_average}:
\begin{align}
\hat{P}_L (k)\equiv\frac{(2L+1)}{I_{22}}\int_{\hat{\bk}} F_L (\bk) F_0 (-\bk)\,,
\label{eq:pl_estimator}
\end{align}
where 
\begin{align}
F_L (\bk)&\equiv\int_{\bx}W(\bx)e^{-i\bk\cdot\bx} \mathcal{L}_L(\hat{\bk}\cdot\hat{\bx})\delta(\bx)\,.
\label{eq:fl_estimator}
\end{align}

To compare with full-sky mocks under Newtonian RSD, the theoretical modeling of galaxy fluctuation $\delta(\bx)$ needs to include the canonical Newtonian RSD term, the angular kernel of which is given in Eq.~\ref{eq:kernel-RSD}. We need to include an additional velocity contribution scaled with $v/x$, which is generated from the Jacobian associated with the change of coordinates caused by Newtonian RSD \cite{98Szalay_RSD_correlation_WA,08Papai_WA,15YooWA,10Raccanelli_WA_sim}. This velocity term is usually ignored for mocks or surveys with small angular coverage \cite{19BeutlerMutipolePk,21BeutlerWAWindow,19Castorina_fnl_quasar}, but it must be included in the full-sky limit \cite{10Raccanelli_WA_sim,24Benabou_WA}. The angular kernel corresponding to this velocity term is
\begin{align}
\Delta^{{\rm v}}_{\ell
}(k,x)=-\frac{\alpha(x)}{x}\frac{f(x)}{k}D(x)j_{\ell}'(kx)\,
\label{eq:alpha-kernel}
\end{align}
where $\alpha(x)\equiv \partial[\ln (x^2 R(x))]/\partial \ln x$, and it becomes $\alpha=2$ for a uniform radial window in the redshift bin \cite{15YooWA,18CastorinaWA}.

The mocks in Ref.~\cite{24Benabou_WA} are generated with $f=0.71$ and $b_1=1.455$, and there is no redshift evolution within the redshift bin. All calculations assume a best-fit Planck 2018 cosmology~\cite{18Planck_Parameter}. The results for the mean of the multipole measurements from the 10,000 mocks with $L$ from 0 to 4 are shown in Fig.~\ref{fig:multipole-sims}. We see that our theory predictions agree exceptionally well with the measured multipoles at the largest scales. The disagreements at the small scales ($k\gtrsim 0.06\,h/{\rm Mpc}$) are due to the voxel windows used in the simulations \cite{24Benabou_WA} which we do not include in our modeling.

\section{Gaussian Covariance}\label{sec:covariance}

We next express the Gaussian covariance of PSM in terms of the gSFB power spectrum. Compared to Ref.~\cite{20Wadekar_covariance}, which had the state-of-art results for analytical Gaussian covariance of the PSM, we will calculate the exact covariance without resorting to any plane-parallel or LOS approximation. The results in Ref.~\cite{20Wadekar_covariance} are only applicable in the flat sky limit and on scales where the window function is subdominant, while our expression is exact and can be applied at the largest scales in particular for measuring PNG and GR effects. Using the Yamamoto estimator of Eq.~\ref{eq:pl_estimator}, the Gaussian covariance reads \cite{20Wadekar_covariance}:
\begin{widetext}
\begin{align}
&\textbf{C}^\textup{G}_{L_1L_2} (k_1,k_2)=\frac{(2L_1+1)(2L_2+1)}{I_{22}^2}\bigg[\int_{\hat{\bk}_1,\hat{\bk}_2} \langle F_{L_1}(\bk_1) F_{0}(-\bk_1) F_{L_2}(\bk_2) F_{0}(-\bk_2) \rangle \bigg] - \langle \widehat{P}_{L_1} (k_1) \rangle\ \langle \widehat{P}_{L_2} (k_2) \rangle\,.
\label{eq:cov-definition}
\end{align}
Using Wick contractions on the 4-point function and expressing $F_{L}(k)$ with Eq.~\ref{eq:fl_estimator}, we can then follow the same procedure as the PSM case. A more detailed derivation will be given in Appendix~\ref{sec:PSM_Cov_Deriv}. To simplify the final expression, we further assume the separability of the window function and obtain:
\begin{align}
&\textbf{C}^\textup{G}_{L_1L_2}(k_1,k_2)=(4\pi)^4\frac{(2L_1+1)(2L_2+1)}{I_{22}^2}\sum_{a,b,c,d,\ell_1,\ell_2}i^{-a-c+b+d}(2a+1)\begin{pmatrix}
a & L_1 & b\\
0 & 0 & 0
\end{pmatrix}^2\begin{pmatrix}
c & L_2 & d\\
0 & 0 & 0
\end{pmatrix}^2 \nonumber\\
&\qquad \Bigg[(2c+1) S_{b\ell_1d\ell_2} +\hr{(-1)^{c+d}}(2d+1) S_{b\ell_1c\ell_2}\Bigg] C_{\ell_1}^{ad,{\rm R}}(k_1,k_2)C_{\ell_2}^{bc,{\rm R}}(k_1,k_2)\,,\label{eq:cov-G}
\end{align}
\end{widetext}
where we have defined $S_{abcd}$ as the chain of window function  (similar to Ref.~\cite{21Gebhardt_SuperFab}):
\begin{align}
S_{abcd}\equiv\sum_{m_a,m_b,m_c,m_d}M^{am_a}_{bm_b}M^{bm_b}_{cm_c}M^{cm_c}_{dm_d}M^{dm_d}_{am_a}\,,
\label{eq:covariance-window-mixing}
\end{align}
and $M_{am_a}^{bm_b}$ as a double spherical harmonic transform of the angular window $M(\hat{x})$:
\begin{equation}
 M_{am_a}^{bm_b}\equiv \int_{\hat{\bx}}M(\hat{\bx})Y_{am_a}(\hat{\bx})Y^*_{bm_b}(\hat{\bx})\,.
\label{eq:second-harmonic-window}   
\end{equation}

Under the separable window assumption, Eq.~\ref{eq:cov-G} expressed the PSM covariance as the sum of the product of the radially-convolved gSFB power spectrum. The formula is feasible for numerical computation due to the triangle conditions of the Wigner-3j symbol and $L_1,L_2$ taking only from 0 to 4. In addition, we usually only need to compute covariance once for the cosmological analysis. 

\begin{figure}[bpt]
\centering
\includegraphics[width=0.4\textwidth]{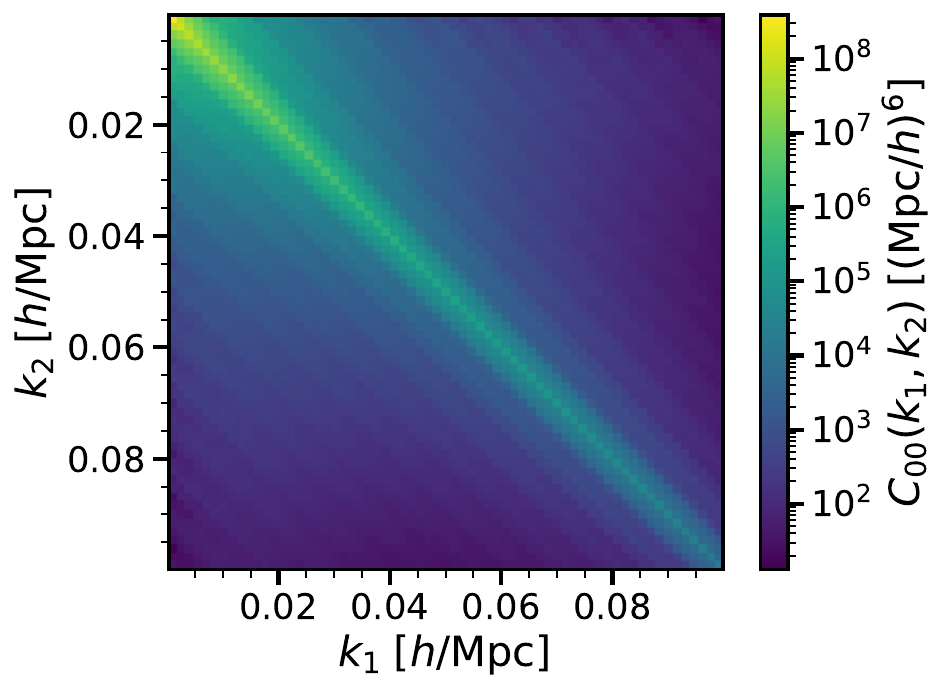}
\caption{The analytical Gaussian covariance for the PS monopole $\hat{P}_0(k)$ evaluated for $z=0.2$ to $0.5$.}
\label{fig:Gaussian-Cov}
\end{figure}

Under the full-sky window, the Gaussian auto-covariance of the monopole $\textbf{C}^\textup{G}_{00}(k_1,k_2)$ simplifies to:
\begin{align}
\textbf{C}^\textup{G}_{00} (k_1,k_2)&=\frac{(4\pi)^4}{I_{22}^2}\sum_{b}2(2b+1)\left[C_{b}^{{\rm R}}(k_1,k_2)\right]^2\,. \label{eq:cov-00}
\end{align}
Under the same set-up as our log-normal mocks with full-sky window, we plot an example of the analytical Gaussian covariance matrix using Eqs.~\ref{eq:gSFB-kernel_form} and \ref{eq:cov-00} in Fig.~\ref{fig:Gaussian-Cov}, which shows prominent off-diagonal components with $k\lessapprox 0.02 h/{\rm Mpc}$.

\begin{figure}[bpt]
\centering
\includegraphics[width=0.375\textwidth]{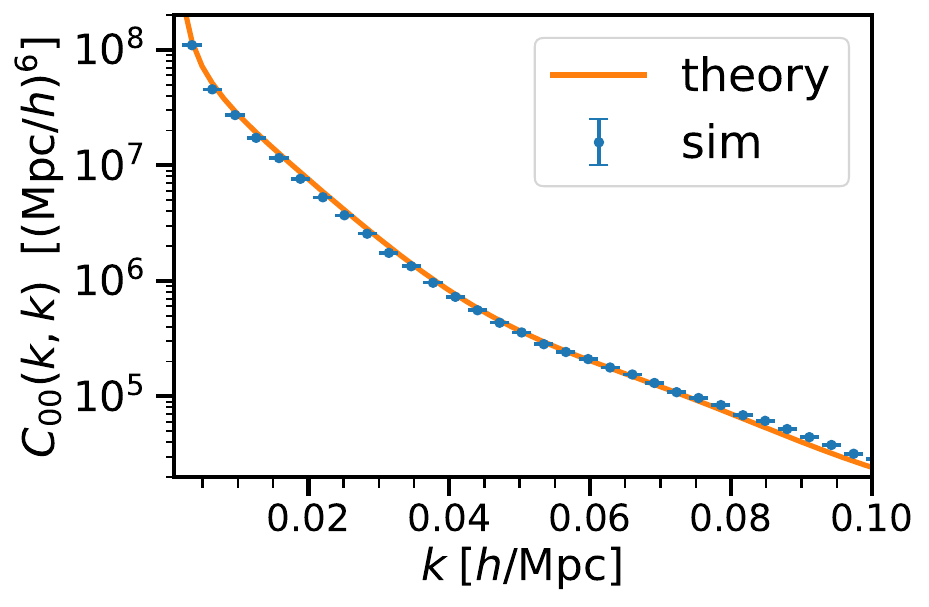}
\caption{Diagonal elements of the auto-covariance for the monopole $\hat{P}_0(k)$ obtained from analytical calculations (orange line, including shot noise) and log-normal mocks (blue points).}
\label{fig:Cov_valid}
\end{figure}

Eq.~\ref{eq:cov-G} gives the Gaussian covariance for a continuous density field. However, real surveys measure discrete objects (galaxies), which produces a non-negligible shot noise contribution \cite{94FKP}. To extend Eq.~\ref{eq:cov-G} for discrete objects, we can include the shot noise to the gSFB power spectrum following Ref.~\cite{21Gebhardt_SuperFab}, which will then allow validation of analytical covariance with covariance estimated from mocks. We compare the diagonal elements of the auto-covariance for the PS monopole obtained from analytical calculations and log-normal mocks in Fig.~\ref{fig:Cov_valid}, and they show good agreement.

Even though the mock-based method of estimating covariance can also include WA effects, it would be computationally expensive to generate a large number of full-sky mocks to achieve the desired accuracy in the covariance. It is also challenging to incorporate GR effects into the mocks, which will require ray-tracing of N-body simulations \cite{17Borzyszkowski_LIGER,20Leopori_lensing_ray_tracing} (or even GR N-body simulations \cite{16Adamek_Gevolution,23Bonvin_dipole_cross_correlation}) of large volumes, so analytical covariance is preferred at large scales. Our result Eq.~\ref{eq:cov-G} lays out the foundation for computing exact analytical covariance at the largest scales for upcoming LSS surveys. 

\section{Conclusion and Discussion}\label{sec:conclusion}

Through Eqs.~\ref{eq:PSM-SFB} and \ref{eq:cov-G}, the main results of this work, we have expressed the PSM and its covariance in terms of the generalized SFB power spectrum. The SFB basis naturally incorporates the WA effect present in the PSM without perturbative expansion of galaxy separation. It also provides a new way to do window convolution of the PSM through the SFB basis, without using the effective redshift approximation \cite{19Castorina_fnl_quasar,22CatorinaGR-P}, which will potentially allow the use of large redshift bin(s) for cosmology with PSM.

Since the mappings in Eqs.~\ref{eq:PSM-SFB} and \ref{eq:cov-G} are independent of cosmological models, we can add GR effects by including additional angular kernels $\Delta_L(k,z)$ \cite{11Bonvin_GR,13DiDio_classgal}. In fact, the GR Doppler effect, which has the same form as Eq.~\ref{eq:alpha-kernel} (just with different coefficients), has already been included in this work. The SFB-to-PSM mapping offers another non-perturbative method for modeling WA and GR effects on PSM in addition to the correlation function approach outlined in Ref.~\cite{22CatorinaGR-P}. 

We are now able to validate the accuracy of the numerical codes for GR effects by comparing results from two different approaches, which will be important for upcoming LSS surveys with the aim to measure $\sigma(\fNL)\sim O(1)$ \cite{23Foglieni}. The natural next step for our work will be to add the remaining GR terms to the gSFB power spectrum, which will enable the computation of the PSM covariance with GR effects, and allow us to forecast the measurability of GR effects and assess their impact on $\fNL$ constraints in the Fourier space for future surveys.

Our result Eq.~\ref{eq:PSM-SFB} suggests that PSM serves as a natural compression of the SFB power spectrum. It will be interesting to study the loss of information caused by such compression. Yamamoto-like estimators assume a pair LOS, which can cause information loss for galaxies with large angular separations. In contrast, SFB allows for a different LOS for every galaxy and fully preserves the redshift evolution, serving as a more optimal statistic. This motivates a search for better compression schemes of the SFB on large scales to potentially improve upon the PSM. 

Alternatively, to avoid the loss of information and the non-linear modeling of SFB power spectrum, one can use a hybrid estimator as suggested in Ref.~\cite{20Wang_Hybrid_Estimator}, that is using the SFB basis on large scales and the Cartesian basis on smaller scales. Employing the hybrid method on upcoming LSS surveys will require further development in the approach, and our exact modeling of PSM at large scales offers the baseline. The SFB-to-PSM mapping can help model the covariance of the hybrid estimator, establish consistency between the two basis, and optimize the transition scale between the two basis to balance constraining power and computational efficiency. Using PSM on large scales still has the advantage of reducing the number of data vectors and allowing the use of a single statistic and a single estimator (the Yamamoto estimator) across all scales, along with simpler joint modeling with bispectrum multipoles.

To apply our formalism for modeling the PSM on realistic data, we have to account for realistic window and efficiently evaluate the generalized SFB power spectrum.  We have so far restricted ourselves to separable windows. For non-separable windows where the redshift distribution is anisotropic, we will need to extend Eqs.~\ref{eq:angular-mask-gSFB} and \ref{eq:cov-G} and evaluate the window convolution of the SFB mode under the generic case as considered in Ref.~\cite{21Gebhardt_SuperFab}. Non-separable windows will cause mixing of both angular and radial modes, which will be significantly more expensive to evaluate. However, the evaluation of the generic window convolution is potentially achievable for the largest scales where there are only a few SFB modes and the effect of a non-separable window is the most prominent.

Throughout the work, we have only considered the linear modeling of PSM. For even multipoles, the wide-angle effects are only important at the largest scales where the linear-order theory is sufficient, so one can use the SFB-to-PSM mapping developed in this work at the largest scales for WA modeling and switch to the standard plane-parallel approximation at the quasi-linear scales for 1-loop calculations. In contrast, the odd multipoles vanish under the plane-parallel approximation for auto-correlation and have to be modeled under the wide-angle regime at all scales. One can then adopt the perturbative WA approach, which only breaks down at the largest scales as shown in Ref.~\cite{24Benabou_WA}, at the quasi-linear scales for 1-loop calculations \cite{23WAGR}. Alternatively, it might be interesting to explore the challenging task of evaluating the SFB power spectrum at the 1-loop order, which can then be mapped to PSM for obtaining both odd and even multipoles beyond the linear theory.

Our mapping Eq.~\ref{eq:PSM-SFB} considered the PSM and the gSFB as theoretical quantities. However, it also applies to them as estimators used on data. It will be interesting to explore the use of the mapping at the estimator level to validate the consistency of the measurements of two statistics. As suggested in Ref.~\cite{18CastorinaWA}, Eq.~\ref{eq:PSM-SFB} can be potentially used to control various systematic effects. In the SFB basis, there is a clear separation between radial and angular scales, so we can remove any systematic-dominated angular modes from the Yamamoto estimator by removing a few SFB modes. 

Furthermore, our work can be extended to the bispectrum. The numerical evaluation of the SFB bispectrum has been recently achieved in Ref.~\cite{23Benabou_SFB}, which will allow for an accurate modeling of WA, GR and window effects present in the bispectrum multipoles on large scales without taking the perturbative approach used in literature \cite{22Pardede_Bi_Window,23Pardede_WA_Bi,23WAGR}.

\section*{Acknowledgements} 
We thank Joshua Benabou, Isabel Sands, Richard Feder, Mike S. Wang, and the SPHEREx cosmology team for useful discussion and feedback. We acknowledge support from the SPHEREx project under a contract from the NASA/GODDARD Space Flight Center to the California Institute of Technology. Part of this work was done at Jet Propulsion Laboratory, California Institute of Technology, under a contract with the National Aeronautics and Space Administration (Contract No. 80NM0018D0004).

\vfill
\appendix

\section{Derivation of SFB-to-PSM Mapping}\label{sec:PSM-SFB_Deriv}

Here we provide a derivation of the SFB-to-PSM mapping in Eq.~\ref{eq:PSM-SFB}. We first expand all the exponential terms in Eq.~\ref{eq:Pl_average} with
\begin{equation}
\label{eq:plane_wave_exp}
    e^{i \bk \cdot \bx} = \sum_\ell i^\ell (2\ell +1) j_\ell(kx) \mathcal{L}_\ell (\hat{\bk} \cdot \hat{\bx})\,,
\end{equation}
and evaluate the $\hat{\bk}$ integral using:
\begin{widetext}
\begin{align}
\int_{\hat{\bk}}\mathcal{L}_{a}(\hat{\bk}\cdot\hat{\bx})\mathcal{L}_{b}(\hat{\bk}\cdot\hat{\bx}')\mathcal{L}_{\ell}(\hat{\bk} \cdot \hat{\bx})=\frac{(4\pi)^2}{2b+1}\begin{pmatrix}
a & \ell & b\\
0 & 0 & 0
\end{pmatrix}^2\sum_{m_b}Y_{bm_b}(\hat{\bx})Y_{bm_b}^*(\hat{\bx}')\,.
\label{eq:L21-k}
\end{align}
This gives us:
\begin{align}
\frac{I_{22}}{(2L+1)}P_{L}(k)&=\sum_{a,b}i^{-a+b}(2a+1)(2b+1)\int_{\bx,\bx'}j_a(kx)j_b(kx')\langle\delta^{{\rm W}}(\bx)\delta^{{\rm W}}(\bx')\rangle\int_{\hat{\bk}}\mathcal{L}_{a}(\hat{\bk}\cdot\hat{\bx})\mathcal{L}_{b}(\hat{\bk}\cdot\hat{\bx}')\mathcal{L}_{L}(\hat{\bk} \cdot \hat{\bx})\nonumber\\
&=(4\pi)^2\sum_{a,b}i^{-a+b}(2a+1)\begin{pmatrix}
a & L & b\\
0 & 0 & 0
\end{pmatrix}^2\sum_{m_b}\langle\int_{\bx}j_a(kx)\delta^{{\rm W}}(\bx)Y_{bm_b}^*(\hat{\bx})\int_{\bx'}j_b(kx')\delta^{{\rm W}}(\bx')Y_{bm_b}(\hat{\bx}')\rangle\,\label{eq:PSM-expand}.
\end{align}
Applying the gSFB decomposition of Eq.~\ref{eq:gsfb_k-to-x} to the windowed field $\delta^{\rm W}(\bx)$ and using the windowed gSFB (pseudo) power spectrum defined in Eq.~\ref{eq:gSFB-W}, Eq.~\ref{eq:PSM-expand} will simplify to Eq.~\ref{eq:PSM-SFB}.

\section{Derivation of PSM Covariance}\label{sec:PSM_Cov_Deriv}
Here we give a derivation of the PSM Gaussian covariance in Eq.~\ref{eq:cov-G}. Starting with the covariance definition in Eq.~\ref{eq:cov-definition}, we split the 4-point function assuming Gaussianity and substitute $F_{L}(k)$ with Eq.~\ref{eq:fl_estimator}:
\begin{align}
&\frac{I_{22}^2}{(2L_1+1)(2L_2+1)}\textbf{C}^\textup{G}_{L_1L_2} (k_1,k_2)=\int_{\hat{\bk}_1,\hat{\bk}_2} \bigg[ \langle F_{L_1}(\bk_1) F_{0}(-\bk_2)\rangle \langle F_{L_2}(\bk_2) F_{0}(-\bk_1) \rangle\nonumber\\
&\qquad + \langle F_{L_1}(\bk_1) F_{L_2}(\hr{\bk_2})\rangle \langle F_{0}(\hr{-\bk_2}) F_{0}(-\bk_1) \rangle\bigg]\nonumber\\
&=\hr{\int_{\hat{\bk}_1,\hat{\bk}_2} \int_{\bx_1,\bx'_1,\bx_2,\bx'_2} e^{-i\bk_1 \cdot (\bx_1-\bx'_1)}\langle \delta^{\rm W}(\bx_1)\delta^{\rm W}(\bx'_2) \rangle\langle \delta^{\rm W}(\bx'_1)\delta^{\rm W}(\bx_2) \rangle   \mathcal{L}_{L_1}(\hat{\bx}_1 \cdot \hat{\bk}_1)} \nonumber\\
&\qquad \hr{\Big[e^{-i\bk_2 \cdot (\bx_2-\bx'_2)} \mathcal{L}_{L_2}(\hat{\bx}_2 \cdot \hat{\bk}_2) +e^{i\bk_2 \cdot (\bx_2-\bx'_2)} \mathcal{L}_{L_2}(\hat{\bx}'_2 \cdot \hat{\bk}_2) \Big ]}\,.
\label{eq:covariance-general}
\end{align}

We consider the first term in the above expression (the $\mathcal{L}_{L_1}(\hat{\bx}_1 \cdot \hat{\bk}_1)\mathcal{L}_{L_2}(\hat{\bx}_2 \cdot \hat{\bk}_2)$ term, denoted as $\textbf{C}^\textup{G-1}_{L_1L_2} (k_1,k_2)$). Expanding all the exponentials with the plane wave expansion Eq.~\ref{eq:plane_wave_exp} and using Eq.~\ref{eq:L21-k} to evaluate the $\hat{\bk}_1$ and $\hat{\bk}_2$ integrals, we obtain
\begin{align}
&\frac{I_{22}^2}{(2L_1+1)(2L_2+1)}\textbf{C}^\textup{G-1}_{L_1L_2} (k_1,k_2)=(4\pi)^4\sum_{a,b,c,d,L_1,L_2}i^{-a+b-c+d}(2a+1)(2c+1)\begin{pmatrix}
a & L_1 & b\\
0 & 0 & 0
\end{pmatrix}^2\begin{pmatrix}
c & L_2 & d\\
0 & 0 & 0
\end{pmatrix}^2\sum_{m_b.m_d}\int_{\bx_1,\bx'_2}\nonumber\\
&\quad\langle \delta^{\rm W}(\bx_1)\delta^{\rm W}(\bx'_2) \rangle j_a(k_1x_1)j_d(k_2x_2')Y^*_{bm_b}(\hat{x}_1)Y_{dm_d}(\hat{x}'_2)\int_{\bx_1',\bx_2}\langle \delta^{\rm W}(\bx'_1)\delta^{\rm W}(\bx_2) \rangle j_b(k_1x'_1)j_c(k_2x_2)Y_{bm_b}(\hat{x}'_1)Y^*_{dm_d}(\hat{x}_2)\nonumber\,.
\end{align}
We can apply the same procedure to the second term containing $\mathcal{L}_{L_1}(\hat{\bx}_1 \cdot \hat{\bk}_1)\mathcal{L}_{L_2}(\hr{\hat{\bx}'_2 \cdot \hat{\bk}_2})$ in Eq.~\ref{eq:covariance-general}. We then apply the gSFB decomposition of Eq.~\ref{eq:gsfb_k-to-x} to the windowed field $\delta^{\rm W}(\bx)$ and the Gaussian covariance becomes:
\begin{align}
&\frac{I_{22}^2}{(2L_1+1)(2L_2+1)}\textbf{C}^\textup{G}_{L_1L_2} (k_1,k_2)=(4\pi)^4\sum_{a,b,c,d,L_1,L_2}i^{-a+b-c+d}(2a+1)\begin{pmatrix}
a & L_1 & b\\
0 & 0 & 0
\end{pmatrix}^2\begin{pmatrix}
c & L_2 & d\\
0 & 0 & 0
\end{pmatrix}^2\Bigg[(2c+1)\sum_{m_b,m_d}\nonumber\\
&\quad\langle \delta_{bm_b}^{a,{\rm W}}(k_1)\delta_{dm_d}^{d*,{\rm W}}(k_2) \rangle \langle \delta_{bm_b}^{b*,{\rm W}}(k_1)\delta_{dm_d}^{c,{\rm W}}(k_2)\rangle+(2d+1)\hr{(-1)^{c+d}}\sum_{m_b,m_c}\langle \delta_{bm_b}^{a,{\rm W}}(k_1)\delta_{cm_c}^{d*,{\rm W}}(k_2) \rangle \langle \delta_{bm_b}^{b*,{\rm W}}(k_1)\delta_{cm_c}^{c,{\rm W}}(k_2)\rangle\Bigg]\label{eq:cov-G-gSFB}\,,
\end{align}
where we have expressed the PSM Gaussian covariance in terms of the gSFB modes.

We now assume the separability of the window function into the angular and radial parts $W(\bx)=R(x)M(\hat{\bx})$. We consider the impact of the angular mask on the gSFB mode, and we find
\begin{align}
\delta_{\ell m}^{a,{\rm W}}(k)=\int_{x}
j_L(kx)\int_{\hat{\bx}}Y^*_{\ell m}(\hat{\bx})M(\hat{\bx})\sum_{\ell'm'}Y_{\ell' m'}(\hat{\bx})\delta^{\rm R}_{\ell'm'}(x)=\sum_{\ell'm'}M^{\ell m}_{\ell' m'} \delta_{\ell' m'}^{a,{\rm R}}(k),
\label{eq:angular_window_convolv}
\end{align}
where the superscript R indicates the presence of only radial selection without any angular mask, and $M^{\ell m}_{\ell' m'}$ is the double spherical harmonic transform of the angular mask defined in Eq.~\ref{eq:second-harmonic-window}. Therefore,
\begin{align}
\langle \delta_{bm_b}^{x,{\rm W}}(k_1)\delta_{dm_d}^{y*,{\rm W}}(k_2) \rangle=\sum_{\ell,m_{\ell}}M_{\ell m_{\ell}}^{bm_b}\sum_{u,m_{u}}M^{um_{u}}_{dm_d}\langle \delta_{\ell m_\ell}^{x,{\rm W}}(k_1)\delta_{um_u}^{y*,{\rm W}}(k_2) \rangle=\sum_{\ell,m_{\ell}}M_{\ell m_{\ell}}^{bm_b}M^{\ell m_{\ell}}_{dm_d}C_{\ell}^{xy,{\rm R}}(k_1,k_2)\,,\label{eq:angular_convolv_gSFB}
\end{align}
where we have used Eq.~\ref{eq:gSFB-def} that defines the gSFB power spectrum under azimuthal symmetry.

We now use the above Eq.~\ref{eq:angular_convolv_gSFB} to simplify the following expression contained in the first term of Eq.~\ref{eq:cov-G-gSFB}:
\begin{align}
&\sum_{m_b,m_d}\langle \delta_{bm_b}^{a,{\rm W}}(k_1)\delta_{dm_d}^{d*,{\rm W}}(k_2) \rangle \langle \delta_{bm_b}^{b*,{\rm W}}(k_1)\delta_{dm_d}^{c,{\rm W}}(k_2)\rangle\nonumber\\
=&\sum_{\ell_1,\ell_2,m_b,m_d,m_1,m_2}M_{\ell_1m_1}^{bm_b}M^{\ell_1m_1}_{dm_d}C_{\ell_1}^{ad,{\rm R}}(k_1,k_2)M^{\ell_2m_2}_{bm_b}M_{\ell_2m_2}^{dm_d} C_{\ell_2}^{bc,{\rm R}}(k_1,k_2)\nonumber\\
=&\sum_{\ell_1,\ell_2}S_{b\ell_1d\ell_2}C_{\ell_1}^{ad,{\rm R}}(k_1,k_2)C_{\ell_2}^{bc,{\rm R}}(k_1,k_2)\,,
\end{align}
where $S_{b\ell_1d\ell_2}$ is the chain of window defined in Eq.~\ref{eq:covariance-window-mixing}. 
One can similarly simplify the second term of Eq.~\ref{eq:cov-G-gSFB}, and the covariance expression in Eq.~\ref{eq:cov-G-gSFB} will reduce to Eq.~\ref{eq:cov-G} under any separable window.

\end{widetext}

\bibliography{refs}

\end{document}